Article

# Impact of pre-patterned structures on features of Laser Induced Periodic Surface structures

Stella Maragkaki [1]*, Panagiotis C. Lingos [1], George D. Tsibidis [1,2], George Deligeorgis [1,2] and Emmanuel Stratakis [1,2]*

1   Institute of Electronic Structure and Laser, Foundation for Research and Technology—Hellas, 71110 Heraklion, Crete, Greece
2   Department of Physics, University of Crete, 71003 Heraklion, Crete, Greece
*   Correspondence: marag@iesl.forth.gr; stratak@iesl.forth.gr

**Abstract:** The efficiency of light coupling to surface plasmon polariton (SPP) represents a very important issue in plasmonics and laser fabrication of topographies in various solids. To illustrate the role of pre-patterrned surfaces and impact of laser polsarisation in the excitation of electromagnetic modes and periodic pattern formation, Nickel surfaces are irradiated with femtosecond laser pulses of polarisation perpendicular or parallel to the orientation of the pre-pattern ridges. Experimental results indicate that for polarisation parallel to the ridges, laser induced periodic surface structures (LIPSS) are formed perpendicularly to the pre-pattern with a frequency that is independent of the distance between the ridges and periodicities close to the wavelength of the excited SPP. By contrast, for polarisation perpendicular to the pre-pattern, the periodicities of the LIPSS are closely correlated to the distance between the ridges for pre-pattern distance larger than the laser wavelength. The experimental observations are interpreted through a multi-scale physical model in which the impact of the interference of the electromagnetic modes is revealed.

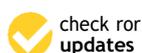

## 1. Introduction

In the past decades, femtosecond laser pulses have become an effective tool in laser processing due to their capacity in precise materials' micro/nano-fabrication with an remarkable contribution to advances in science, technology and industry [1–6]. One particular type of patterns comprising the Laser Induced Periodic Surface Structures (LIPSS) is the most common type of topography that can be fabricated via irradiation of solids' surfaces by Gaussian beams [1,7]. Various laser parameters are usually controlled to tailor not only the LIPSS sizes and periodicities but also the complexity of the induced surface patterns. The energy dose, fluence and laser polarisation state [1,8,9] are among those parameters that are usually modulated to control those features or even increase the complexity of the topographies and therefore allow fabrication of morphologies attained [1].

In principle, LIPSS result from the interference of the incident light and an induced electromagnetic wave that is scattered from the solid surface due to the presence of in-homogeneities or roughness [10]. The formation of the periodic structures is sufficiently described through a detailed investigation of complex multiscale physical mechanisms that include spatially modulated energy absorption from the material, electron excitation, relaxation processes and phase transitions [11–13]. On the other hand, the physical process that leads to the above topographies is based on the concept that upon relaxation a thin molten layer appears on the surface if laser conditions allow a phase transition; due to the spatial temperature gradient in the layer strong instabilities are generated that produce fluid movement and eventually upon resolidification LIPSS are formed [13–18]). Although the importance of the laser parameters as well as the environmental conditions on the LIPSS characteristics is well reported, few studies have been focused on the impact of irradiation of surfaces of enhanced roughness and the role of the excited surface electromagnetic waves on the induced LIPSS topography [10,19,20]. More specifically, given that an increased roughness which can be produced through a multipulse irradiation influences the wavelength of the excited electromagnetic surface waves (i.e. Surface plasmons) and therefore the LIPSS formation and features [13,16,21,22], it is important to investigate the extent to which a pre-patterned sub-micrometer sized surface is capable to guide topography changes. Therefore, elucidation of the impact of the pre-patterned structures on the generation of LIPSS can provide a better insight on the underlying physical mechanisms giving rise to controllability and repeatability of the process in different materials. In a previous study, on rippled structures on pre-patterned surfaces, Miyaji et.al. [23] reported on nanostructure formation on the crests of diamondlike carbon pre-patterned films due to a preferential enhancement of localized electric fields in those regions.

Despite the important results reported in previous works, there are still several unexplored issues that need to be addressed in order to understand thoroughly the process of LIPSS formation upon irradiation of pre-patterned surfaces. As surface plasmon wave propagation wavelength and direction is correlated with the periodicity and orientation, respectively, of the produced LIPSS, a prominent question is whether the pre-pattern interdistance affects their characteristics (i.e. shape and size). Similarly, as electromagnetic effects are strongly related with the spatial distribution of the absorbed energy, a detailed analysis of the role of the pulse-by-pulse topography change in the energy modulation and absorption is very important to determine the process to LIPSS formation. In a recent study by Radko et al [24], it was shown that the predominant mechanism that accounts for LIPSS formation, the excitation of Surface Plasmon Polaritons is dependent on the ridges period, height and number and therefore, electromagnetic effects need to be addressed properly. Similar conclusions had been inferred in other reports with respect to the role of the prepattern features on the LIPSS formation [25]. Furthermore, as the produced hydrothermal waves and resolidification process will provide the features of the surface pattern, it is of importance to investigate how the pre-pattern interacts with the produced melt fluid during the phase transformation and the extent to which the ridges drive the molten material towards forming LIPSS.

Therefore, to fully understand the topography evolution upon irradiation of pre-patterned surfaces and evaluate their impact on LIPSS characteristics, a combined experimental and theoretical approach is presented in this work. To this end, we report on the fs laser induced periodic surface structure formation on pre-patterned Nickel surfaces, featuring different periods and orientation. Furthermore, a detailed analysis of the electromagnetic modes that are developed on the different topographies as a result of the increase of the energy dose is performed to describe the interpulse surface morphology changes.

## 2. Materials and Methods

The formation of the periodic pre-patterns is induced through electron-beam lithography (EBL) using a method presented in Fig. 1.

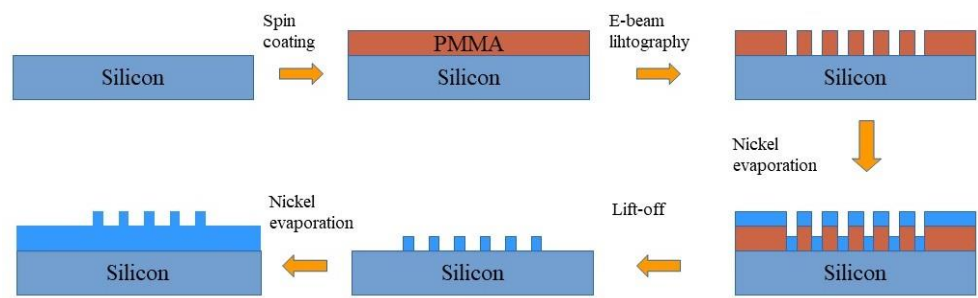

**Figure 1.** Schematic representation of the lift-off fabrication process.

A pristine silicon wafer (100 n-type 1-10 Ω cm) was used as the mechanical substrate. The desired patterns were defined using a Polymethyl methacrylate (PMMA) mask and e-beam lithography (RAITH Quantum combined with a JEOL 7000 F FESEM at 30 KVolts) was used to define a set of periodic structures consisting of lines (50% crest / 50% depression) with a specific periodicity on each site. To ensure pattern fidelity a proximity corrected dose pattern was used. Following PMMA development, 100 nm Nickel was evaporated and lift-off was used to define the Nickel lines. After lift-off, a blanket evaporation of 200 nm Nickel was used to realise the final Nickel structure that is 200 nm Nickel with 100 nm thick Nickel protrusions. The pattern fidelity is ensured by the directionality of the Nickel evaporation. The period was varied by changing the deposition pattern rastered by the electron beam during EBL, optimising the dose for each pattern in order to obtain the required dimensions. The pre-patterned surfaces were subsequently subjected to ultrashort pulsed-irradiation. For this purpose, the fundamental and the second harmonic (1026 nm and 513 nm) of an Yb:KGW laser system, delivering pulses of 170 fs pulse duration at 1 kHz repetition rate were used. Specifically, the Nickel sample is mounted on a xyz translation stage normal to the incident laser beam, which is focused on the sample surface by a lens with 200 mm focal length. The pulse energy is adjusted by means of a variable attenuator formed by a half-waveplate and a polarizing beam splitter. The fluence values were calculated as $\Phi_0 = 2E_P/\pi \omega_0^2$ with $E_P$ the pulse energy and $\omega_0$ the beam radius at $1/e^2$. Following irradiation, the surface morphology was analyzed by means of field-emission scanning electron microscopy. The corresponding LIPSS period was determined via the application of a two dimensional fast fourier transform on the obtained SEM micrographs.

## 3. Results and Discussion

To reveal the role of the interaction of light to the excited surface plasmon waves in the process towards specific topography formation, we analysed the experimental findings for the periodicities of the induced LIPSS in various laser conditions and distances between the periodic pre-pattern arrangement. The LIPSS periods either remain independent of the periods of the pre-patterns or they are influenced by them. The latter case is an indication that the ridges topology itself tailors the interference of the electromagnetic modes that are excited. More specifically, the experimental results at both laser wavelengths (513 nm and 1026 nm) used, reveal the absence of any impact on the LIPSS periodicity. Indeed, as shown in Fig. 2 when the pre-patterned arrangement has a direction parallel to the laser polarisation $E$, the final LIPSS period is constant and independent of the pre-patterned period. The discrepancy between the simulated (see next section) and experimental values is attributed to the precision in the estimation of the experimental conditions, however, a similar constant trend in the evolution is observed in both cases.

By contrast, when the laser polarisation is perpendicular to that of the pre-structures, there is a strong impact of the pre-pattern period on the final spatial topographies. In this case, the LIPSS period follows a distinct behaviour (Fig. 3) that requires further investigation of the underlying physical processes involved. More specifically, for ridge distances of the size or slightly smaller than the laser wavelength, the LIPSS period appears to be close to the SPP wavelength. On the other hand, at larger pre-patterned distances, the LIPSS period is half the pre-pattern distance ($\Lambda_{LIPSS} = \Lambda_{pre-pattern}/2$).

Finally, for pre-patterns spacing which is substantially greater than the laser wavelength (almost twice as much as the laser wavelength) the correlation of the LIPSS period to the pre-pattern distance is provided by the following expression

$$\Lambda_{LIPSS} = \frac{\Lambda_{pre-pattern}}{4} \qquad (1)$$

The above observations are summarised in Table 1. It is evident that while the formation of LIPSS oriented perpendicularly to the periodic arrangement of ridges can be interpreted through the excitation of SPP along the direction of the pre-pattern (see simulated results below), the situation is more complex when the polarisation of the laser beam is perpendicular to the pre-pattern. To simulate the surface modification process, a multiscale description of the underlying physical mechanisms is used aiming also to interpret the interpulse topography changes.

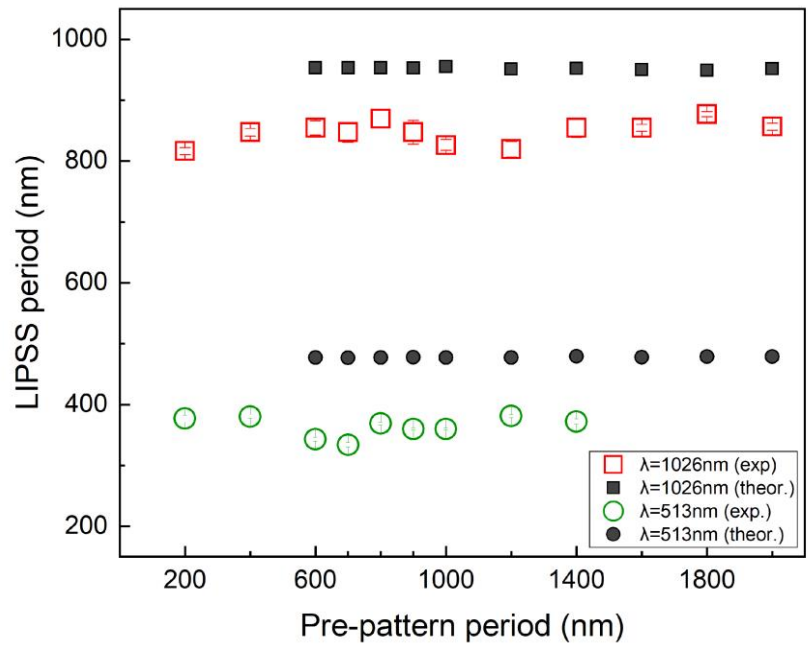

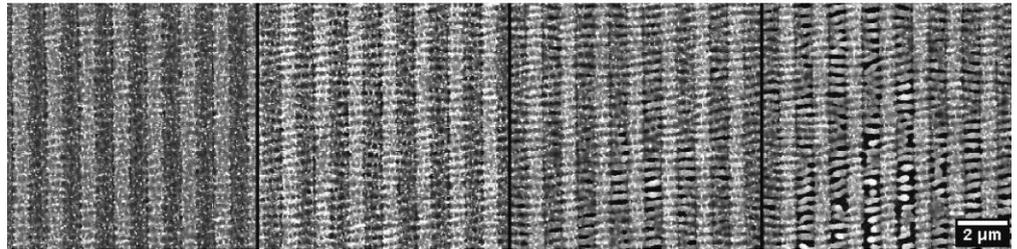

**Figure 2.** LIPSS spatial period as a function of pre-pattern periodicities. LIPSS on Nickel surfaces upon irradiation with 20 laser pulses ( plot) with *polarization parallel to the pre-pattern orientation*. Experimental (*red* empty squares) and theoretical (*black* filled squares) values of the LIPSS periodicity upon irradiation with 1026 nm laser pulses. The experimental (*green* empty dots) and theoretical (*black* filled dots) values of the LIPSS periodicity upon irradiation with 513 nm laser pulses are also presented. The SEM micrographs below represent the evolution of LIPSS on pre-patterns with 1400 nm period upon irradiation with 4, 10, 15 and 20 pulses (from left to right), following irradiation with pulses of wavelength $\lambda$=513 nm and peak fluence 0.12 $J/cm^2$)

### 3.1. Numerical Simulations

Given the experimentally attained distinct changes in the surface topography that depends on the laser polarisation direction, a detailed investigation of the electromagnetic effects is firstly elaborated; more specifically, it is known that scattering of light by a patterned or a corrugated surface is expected to reveal the characteristics of the induced surface waves that operate as precursors of the final topography [12,24,26]. The interference of the incident beam with the surface waves leads to a spatial distribution of the energy that is absorbed from the material which in turn, excites the electron system [13,15,19,27]. A precise evaluation of the electromagnetic effects is performed through the employment of Finite Integration Technique (FIT) algorithms to solve Maxwell's equations using the commercial software CST Studio Suite. FIT [28,29] is a generalized finite difference scheme for the numerical solution of the Maxwell's equations in their integral form (Eq. 2)

| λ (nm) | Λ$_{pre-pattern}$ (nm) | $\frac{\Lambda_{LIPSS}}{\Lambda_{pre-pattern}}$ |
|---|---|---|
| 1026 | < 1200 | independent |
| 1026 | 1200-1600 | 1/2 |
| 1026 | ≥ 1800 | 1/4 |
| 513 | < 600 | independent |
| 513 | 600-900 | 1/2 |
| 513 | ≥ 1000 | 1/4 |

Table 1: The final period of LIPSS as a function of the laser wavelength and the pre-pattern spacing. Final period remains unaffected on the surface topography when the pre-existing period is comparable to the laser wavelength. This table is valid when the laser polarization is perpendicular to the pre-pattern period direction

$$\oint_{\partial A} \mathbf{E} \cdot d\mathbf{s} = - \oiint_A \frac{d\mathbf{B}}{dt} \cdot d\mathbf{A} \tag{2a}$$

$$\oiint_V \mathbf{B} \cdot d\mathbf{A} = 0 \tag{2b}$$

$$\oint_{\partial A} \mathbf{H} \cdot d\mathbf{s} = \oiint_A (\frac{\partial \mathbf{D}}{\partial t} + \mathbf{J}) \cdot d\mathbf{A} \tag{2c}$$

$$\oiint_V \mathbf{D} \cdot d\mathbf{A} = \iiint_V \rho dV \tag{2d}$$

where the field and flux vectors are related to the material equations (Eq. 3),

$$\mathbf{B} = \mu \mathbf{H} + \mathbf{M} \tag{3a}$$
$$\mathbf{D} = \epsilon \mathbf{E} + \mathbf{P} \tag{3b}$$
$$\mathbf{J} = \sigma \mathbf{E} \tag{3c}$$

For a description of the variables in Eq. 2 and Eq. 3, see [28,29].

FIT provides reliable solutions for the entire range of electromagnetic field problems with complex geometries and shares similarities in accuracy with the widely used Finite Difference in the Time-Domain method (FDTD). FIT employs a pair of staggered grids, the primary grid and the dual grid such as the Yee cells in FDTD method. The primary grid composes the entire computational domain as a collection of volume cells $V_i$ ($i = 1..n_V$) surrounded by facets $A_i$ ($i = 1..n_A$) and edges $L_i$ ($i = 1..n_L$). The dual grid is constructed so that each edge of the grid penetrates the surfaces of the other grid and each mesh point of one grid lies at the center of the other grid. Each vector field and flux is converted in integral forms (scalars), the so called state variables. The calculation of material parameters is based on computing the integrals over the surface of the grid cells rather than on the definition of material data on grid points. The discretion form of the Maxwell equation's leads to a complete discrete set of matrix equations known as "Maxwell Grid Equations" representing integrations along edges and over facets of the grid. For a detailed description of FIT-technique, see Ref. [30].

In order to elucidate the formation of LIPSS on a pre-patterned surface irradiated with a laser beam perpendicularly polarised with respect to the orientation of the ridges, a detailed investigation of the electromagnetic intensity distribution over the entire structure is required. Due to the symmetry of the pre-pattern along the $y$ axis, the problem can be reduced from a fully 3D to a quasi-2D focusing on one "unit-cell" of the pre-pattern with period Λ$_{pre-pattern}$. Since the length of the pre-patterned structures in the $y$ direction surpasses the laser beam spot ($\approx 60\mu$m) and the beam spot radius is much greater than the studied pre-patterned periods in the experiments ($\Lambda_{pre-pattern} < 2\mu$m), the laser field is simulated as a plane wave of wavelength $\lambda$=1026 nm or 513 nm, linearly polarized along the $x$ axis (perpendicular to the pre-pattern) or $y$ axis (parallel to the pre-pattern) irradiating the material at normal incidence. The geometry of the problem allows to keep the boundary conditions simple and the computational grid is terminated by two convolutional perfectly matched layers (convPML) in the $z$ direction to avoid non-physical reflections while periodic boundary conditions are used for $x$ and $y$ directions. In order to capture the details of each pre-patterned surface i.e corrugations after irradiation and re-solidification, the grid cell dimensions are selected to be a fraction of the wavelength. In our simulations, we used 50 to 60 grid cells per wavelength applied in all directions. The values of the refractive index $n$ and extinction coefficient $k$ for Nickel are based on fitting experimental data with a Lorentz-Drude model [31]. The pre-patterned structure consists of two 100 nm thick and $\Lambda_{pre-pattern}/4$ wide Nickel protrusions with spacing $\Lambda_{pre-pattern}/2$ between them. It is noted that the irradiated solid is thick enough to be considered optically as an nearly infinite layer (i.e. for the laser wavelengths used in the simulations $\lambda = 1026$ nm and $\lambda = 513$ nm, the skin depth of the Nickel is $\delta \approx 32$ nm and $\delta \approx 27$ nm, respectively). In our simulations, we computed the intensity $I = |\mathbf{E}|^2$, where $\mathbf{E}$ is the complex electric field, normalized by the maximum value $I_0$ emitted by the plane wave source.

A three dimensional Two Temperature Model (TTM) is then, applied to describe the energy transfer from the excited electron system to the lattice subsystem. To evaluate the influence of light on the evolving (i.e. pulse by pulse) topographies, a multi-physics approach is followed to numerically describe energy absorption, excitation, relaxation and phase transitions that eventually lead to an evolution of the irradiated pattern. [15,32,33]. Due to fact that the lattice temperatures that are attained from the exposure of the solid to extreme laser conditions exceed the melting point of the material, a detailed investigation of the dynamics of the produced fluid is conducted to determine the morphological features of the final topography. A thorough description of the multiscale methodology used to simulate surface modification has been presented in several previous reports [8,13,15,18]. Below, results of the modelling approach are summarised that demonstrate the role of the direction of the polarization beam, the influence of the characteristics of the electromagnetic modes and the impact of the pre-patterned period in the periodicity of the induced LIPSS. Simulations have been performed with pulse duration equal to $\tau_p = 170$ fs and peak fluence equal to 0.12 J/cm$^2$ and for different sizes of the pre-pattern periods $\Lambda_{pre-pattern}$ =700 nm, 1400 nm for $\lambda$ =513 nm and $\Lambda_{pre-pattern}$=700 nm, 800 nm, 1900 nm and 1300 nm for $\lambda$ =1026 nm ).

### 3.2. Laser polarization parallel to the pre-pattern orientation

The predominant electrodynamic process that accounts for the formation of LSFL structures is the excitation of surface plasmons waves (SPW) and interference with the incident beam [26]. In previous reports, it has been shown that $\Lambda_{LIPSS}$ is dependent on the number of pulses while the SPW propagates perpendicularly to the laser polarisation [15,22]. Simulations indicate that the SPW propagation is along the direction of laser polarisation while the pre-pattern period does not play any role in the frequency of the waves. Similarly, the propagation of the thermo-capillary waves that are produced as a result of the absorbed energy yields LIPSS of constant periodicity. Theoretical predictions shown in Fig. 4 illustrate an upper view and side view, respectively, of the induced pattern following irradiation with fifteen pulses. These simulations indicate that ripples are forming both on the pre-pattern and the region between the ridges which is confirmed from experimental observations (see Fig 2) for various number of pulses. The periodicity of the induced LIPSS is predicted to be independent of the $\Lambda_{pre-pattern}$ which is explained by the fact that the electromagnetic waves that propagate along the pre-patterned surface are not influenced by the distance of the ridges as they propagate in a direction parallel the patterns; hence, the resultant periodic energy modulation will produce thermocapillary waves propagating in a direction that is not constrained by the pre-pattern orientation.

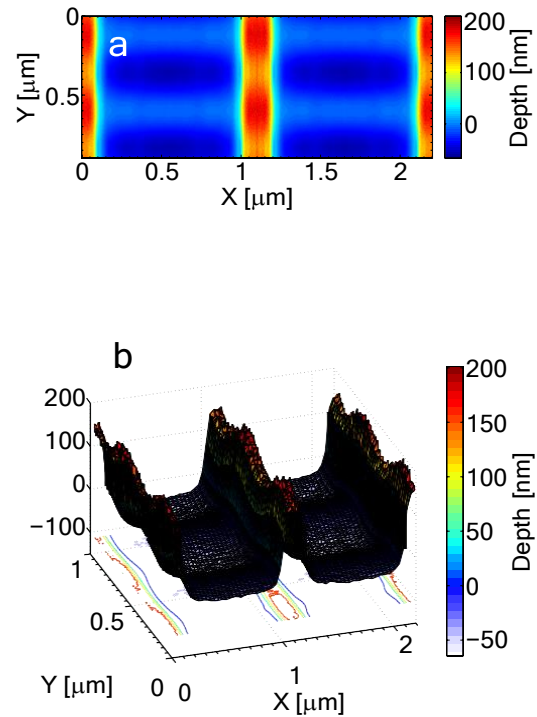

**Figure 4.** Topography following irradiation of Nickel with fifteen laser pulses at 1026 nm (a) Upper view, (b) Side view. Polarisation is along the Y-axis.

### 3.3. Laser polarization perpendicular to the pre-pattern orientation

By contrast, a more complex scenario occurs when the laser polarisation is perpendicular to the ridges orientation.

To examine whether the electromagnetic factor influences the spatial distribution of the energy accumulation on the pre-patterned surface for different $\Lambda_{pre-pattern}$, numerical calculations have been performed to evaluate the impact of the pre-pattern in the energy absorption. For the sake of simplicity, simulations are restricted to the investigation of the topography interpulse changes for four pulses that is expected to capture the pattern periodicities of the produced morphology. In Fig. 5 and Fig. 6, the evolution of topographies are illustrated as a function of $\Lambda_{pre-pattern}$ and $\lambda$. Electromagnetic simulations show the spatial distribution of the electromagnetic field on a pre-pattern surface in one period. Theoretical results indicate that assuming energy absorption for pre-patterns periodicities larger than twice the laser wavelength, two hills are formed inside the valley that are $\Lambda_{pre-pattern}/4$ apart (Fig. 5 (e-h) and Fig. 5 (m-p), which is explained by the interference of the electromagnetic waves that are excited on the ridges. By contrast, at smaller inter-pattern periodicities, one hill is produced at the centre of the valley (Fig. 5 (a-d) and Fig. 5 (i-l)). which is again pertinent to the behaviour of the excited electromagnetic modes. Furthermore, for pre-pattern distances of the size of the laser wavelength, topographies of periodicity *independent* of $\Lambda_{pre-pattern}$ are fabricated and periodic structures of the size of the excited SPP is produced (Fig. 6). Such periodicities have been observed experimentally and predicted from theoretical models in various other previous reports (see [1] and references therein).

The above investigation and results define a methodology to enhance the efficiency of coupling of the electromagnetic modes that are excited with the incident beam which

8 of 12this is extrademonstrates, conclusively, the impact of the geometrical characteristics of the pre-patterned surface on the periodicities and orientation of the induced LIPSS. More specifically, the constructive interference that is caused in the case of irradiation with polarised beam perpendicularly to the ridge orientation could be used to produce LIPSS of variable periodicity by modulating the distance between the pre-patterns. The capability to control and optimise the periodicity of a rippled pattern and fabrication nano/micro size topographies by regulating the pattern features and laser polarisation can provide unique opportunities for promising biosensing or plasmonic applications.

## 4. Conclusions

In summary, we have investigated experimentally and numerically the efficiency of SPP excitation on periodic sets of ridges of Nickel surfaces. The laser parameters, such as the polarisation and wavelength and the pre-pattern features (i.e. orientation with respect to the laser polarization and distance between ridges) were varied and LIPSS were fabricated with periodicities ranging between $\sim \Lambda_{pre-pattern}/4$ and $\lambda_L$ which are dependent on the polarisation of the incident beam and the periodicity of the ridge arrangement. The detailed analysis of the LIPSS formation in different conditions and ridge distances through the evaluation of the electromagnetic effects and the incorporation of a multiscale physical modelling approach and experimental validation can set the basis for control of the parameters to fabricate patterns with desired properties for a wide range of optoelectronic applications.

**Author Contributions:** Conceptualization, S. Maragkaki ; methodology, S. Maragkaki., G.Deligeorgis; Theoretical modelling and software, P. Lingos, GD. Tsibidis; writing—original draft preparation, S.Maragkaki; writing—review and editing, S.Maragkaki, P.Lingos, GD. Tsibidis, G.Deligeorgis, E.Stratakis; All authors have read and agreed to the published version of the manuscript.

**Funding:** This research was funded by the following funding schemes: (1) Ruhr University Research School PLUS, funded by Germany's Excellence Initiative [DFG GSC 98/3], (2) the European Union's Horizon 2020 research and innovation program through the project BioCombs4Nanofibres (Grant Ggreement No. 862016). (3) HELLAS-CH project (MIS 5002735), implemented under the "Action for Strengthening Research and Innovation Infrastructures,". (4) Competitiveness, Entrepreneurship & Innovation" (EPAnEK), Nanoroll (MIS 5048498), ESPA 2014–2020 Program.

**Acknowledgments:** The authors would like to acknowledge A. Lemonis for the technical support and A. Manousaki for the SEM characterization

**Conflicts of Interest:** The authors declare no conflict of interest.

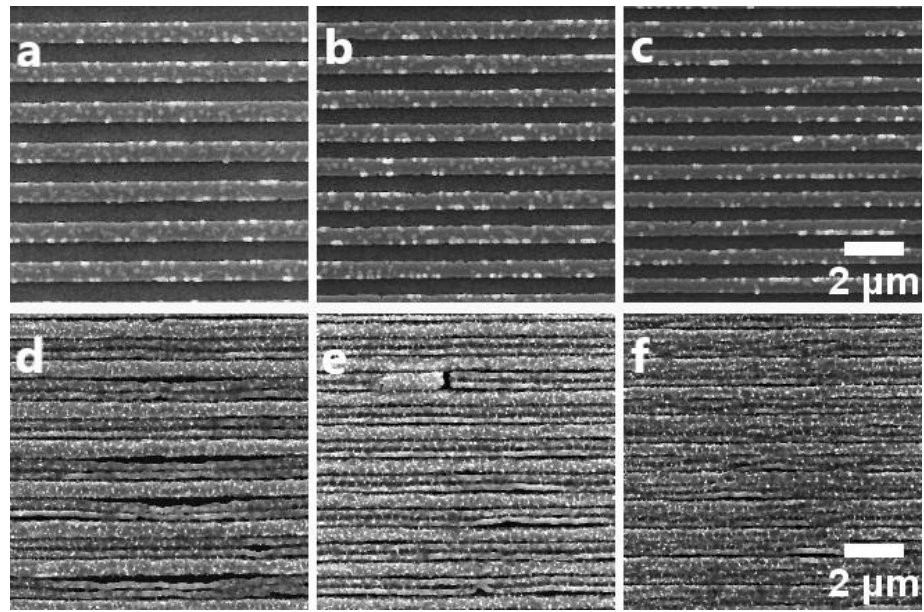

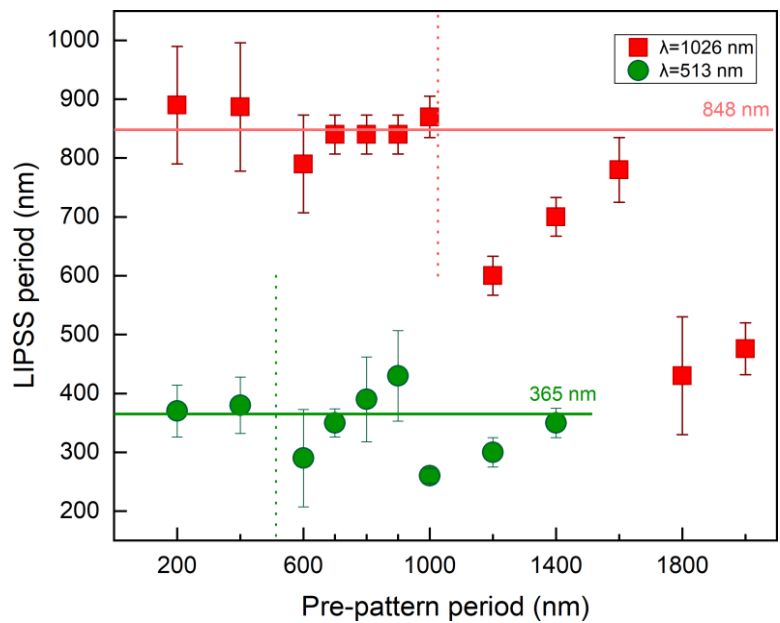

**Figure 3.** LIPSS on Nickel surfaces with different pre-pattern periodicities and a *pre-pattern orientation perpendicular to the laser polarization*. The SEM micrographs represent a comparison of non-irradiated pre-pattern structures (a, b, c) with periods 1400, 1200 and 1000nm respectively and laser-induced ripples on the same pre-pattern structures with LIPSS period 350, 300 and 250nm respectively (d, e, f). The upper row (a,b,c) show the pre-pattern period and orientation. Laser parameters are the same for d, e and f. Femtosecond laser pulses with 513 nm laser wavelength, 25 laser pulses per spot and peak fluence at 0.12 J/cm$^2$ Irradiation with IR (filled squares) and visible (filled dots) fs laser pulses, with laser polarization perpendicular to the pre-pattern orientation. The vertical dotted lines represent the size of the laser wavelength. The horizontal lines correspond to the average LIPSS period when laser polarization is parallel to the pre-structures (see Fig. 2). Here, LIPSS spatial period is independent of the pre-pattern spacing when the spacing is equal to or smaller than the laser wavelength. In contrast, at higher pre-pattern spacing, the pre-pattern has a great impact on the final LIPSS formation.

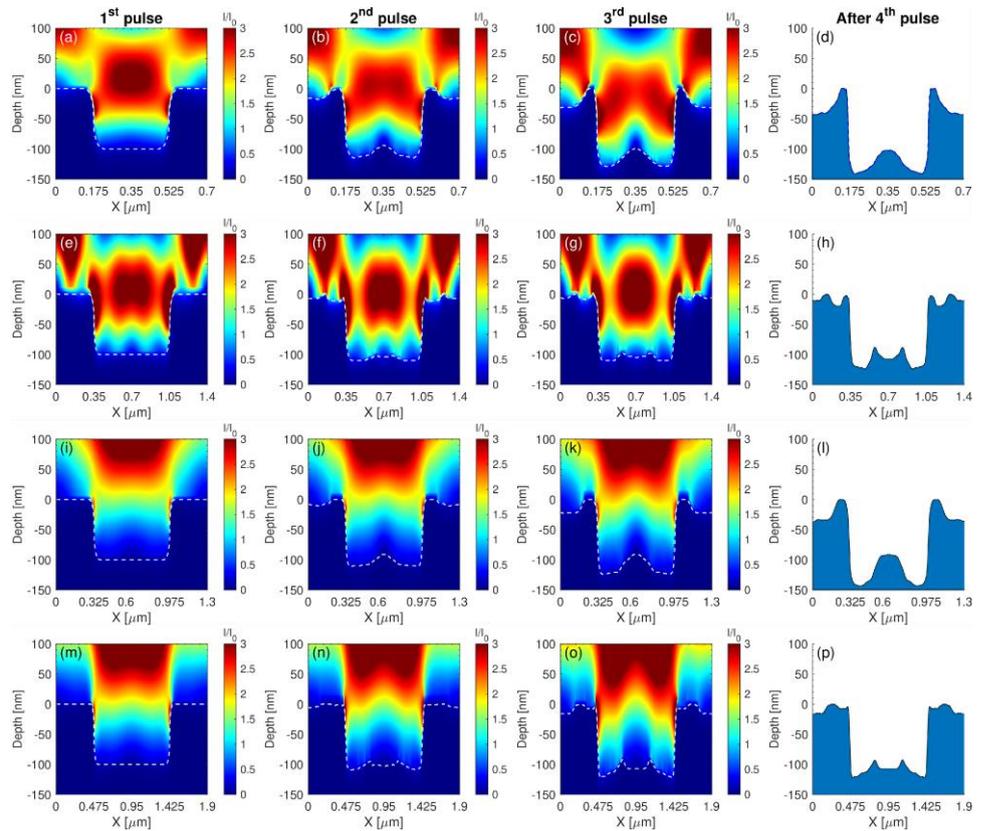

**Figure 5.** Spatial profile of the intensity distribution for different pre-pattern distances $\Lambda_{pre-pattern}$ and laser wavelengths $\lambda$ on the $x$-$z$ plane. In these images the pre-pattern topography evolution upon successive irradiation is also captured (*white* dashed lines). The first, second and third column represent the intensity distribution during the first, second and third laser pulse, respectively, while the final LIPSS topography after four pulses is depicted in the fourth column (side view of the morphology). (a-d) $\lambda$ = 513 nm for $\Lambda_{pre-pattern}$ =700 nm, (e-h) $\lambda$ = 513 nm for $\Lambda_{pre-pattern}$ =1400 nm, (i-l) $\lambda$ = 1026 nm for $\Lambda_{pre-pattern}$ =1300 nm, (m-p) $\lambda$ = 1026 nm for $\Lambda_{pre-pattern}$ =1900 nm. The *white* dashed line indicates the surface boundary.

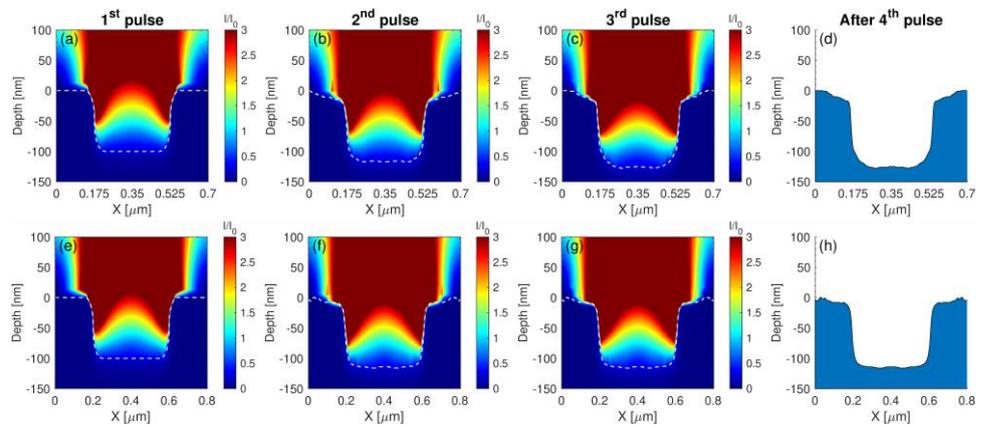

**Figure 6.** Spatial profile of the intensity distribution for different pre-pattern distances (a-c) $\Lambda_{pre-pattern}$ = 700 nm and (e-g) $\Lambda_{pre-pattern}$ = 800 nm during three successive irradiations with laser pulses of $\lambda$ = 1026 nm and the corresponding final topographies (d) and (f) after the fourth pulse. The *white* dashed line indicates the surface boundary.